\documentclass{article}

\usepackage{PRIMEarxiv}

\usepackage[utf8]{inputenc} 
\usepackage[T1]{fontenc}    
\usepackage{hyperref}
\usepackage{url}            
\usepackage{booktabs}       
\usepackage{bm}

\usepackage{soul}
\usepackage{stmaryrd}
\usepackage{amsfonts}       
\usepackage{amsmath}
\usepackage{amssymb}
\usepackage{authblk}
\usepackage{nicefrac}       
\usepackage{microtype}
\usepackage{cleveref}
\usepackage{xcolor}
\usepackage{algpseudocode}
\usepackage{algorithm}
\usepackage{color,soul}
\usepackage{dsfont}
\usepackage{breqn}
\usepackage{caption}
\usepackage{lipsum}
\usepackage{fancyhdr}       
\usepackage{graphicx}       
\graphicspath{{media/}}     
\usepackage{enumitem}
\makeatletter
\newcommand{\printfnsymbol}[1]{%
  \textsuperscript{\@fnsymbol{#1}}%
}

\makeatother

\title{TreeTOp: Topology Optimization using Constructive Solid Geometry Trees }

\author[1]{Rahul Kumar Padhy}
\author[1]{Pramod Thombre}
\author[1,*]{Krishnan Suresh}
\author[2]{Aaditya Chandrasekhar}

\affil[1]{Department of Mechanical Engineering, University of Wisconsin-Madison, Madison, WI, USA 
\authorcr
   \{\tt rkpadhy, pthombre, ksuresh\}@wisc.edu}

\affil[2]{Department of Mechanical Engineering, Northwestern University, Evanston, IL, USA 
\authorcr
    cs.aaditya@gmail.com
  }

\affil[*]{Corresponding author}

\begin{document}
\maketitle

\begin{abstract}

Feature-mapping methods for topology optimization (FMTO) facilitate direct geometry extraction by leveraging high-level geometric descriptions of the designs. However, FMTO often relies solely on Boolean unions, which can restrict the design space. This work proposes an FMTO framework leveraging an expanded set of Boolean operations, namely, union, intersection, and subtraction. The optimization process entails determining the primitives and the optimal Boolean operation tree. In particular, the framework leverages a recently proposed unified Boolean operation approach. This approach presents a continuous and differentiable function that interpolates the Boolean operations, enabling gradient-based optimization. The proposed methodology is agnostic to the specific primitive parametrization and is showcased through various numerical examples.

\begin{figure}[H]
 	\begin{center}
		\includegraphics[scale=0.7,trim={0 0 0 0},clip]{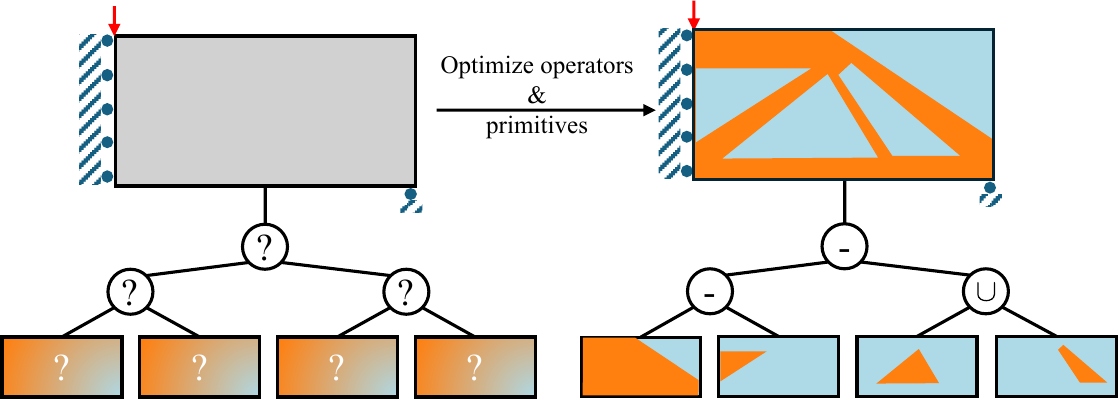}
 		\caption*{Graphical abstract: We perform topology optimization by optimizing the parameters of primitives and the Boolean operations.}
        \label{fig:graphical_abstract}
	\end{center}
 \end{figure}
\end{abstract}


\section{Introduction}
\label{sec:intro}
Topology optimization (TO) has emerged as a powerful computational tool for achieving optimal material distribution within a design domain, subject to constraints \cite{bendsoe2013topology, sigmund2013topology}. The maturity of TO methods is evident in their widespread industrial application, particularly facilitated by commercial software.

\begin{figure}[H]
 	\begin{center}
		\includegraphics[scale=0.8,trim={0 0 0 0},clip]{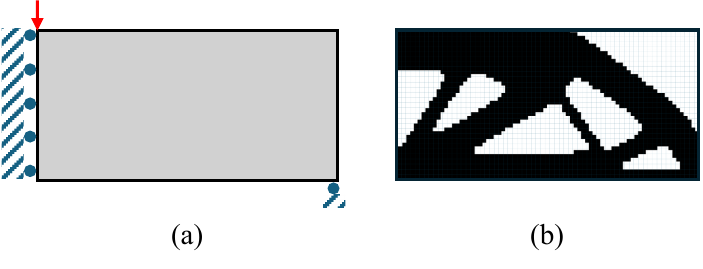}
 		\caption{(a) Design domain and boundary conditions. (b) An optimized design obtained through density-based TO. }
        \label{fig:TO_approaches}
	\end{center}
 \end{figure}

While various TO methods exist, density-based approaches are widely adopted \cite{sigmund2013topology}. Popular density-based approaches discretize the design space into finite elements and optimize a fictitious material density within each element to generate organic, free-form designs \cite{wein2020reviewFeatureMapping}. Consider, for instance, the design domain and boundary conditions in \cref{fig:TO_approaches}(a). An optimal design that maximizes stiffness, subject to a volume constraint, using density-based TO is illustrated in \cref{fig:TO_approaches}(b). While offering design freedom, the resulting designs can be challenging to interpret \cite{wein2020reviewFeatureMapping, ren2021csg} and modify \cite{geometry2022neural}. Furthermore, the density-based TO designs often require extensive post-processing \cite{subedi2020review} that leads to a deviation between the final design and the optimal solution \cite{zhu2020design, zhang2020adaptive}. This deviation becomes more pronounced when the structure is manufactured using stock material, with components of fixed shape but variable dimensions \cite{zhang2020adaptive}.

\begin{figure}[H]
 	\begin{center}
		\includegraphics[scale=0.5,trim={0 0 0 0},clip]{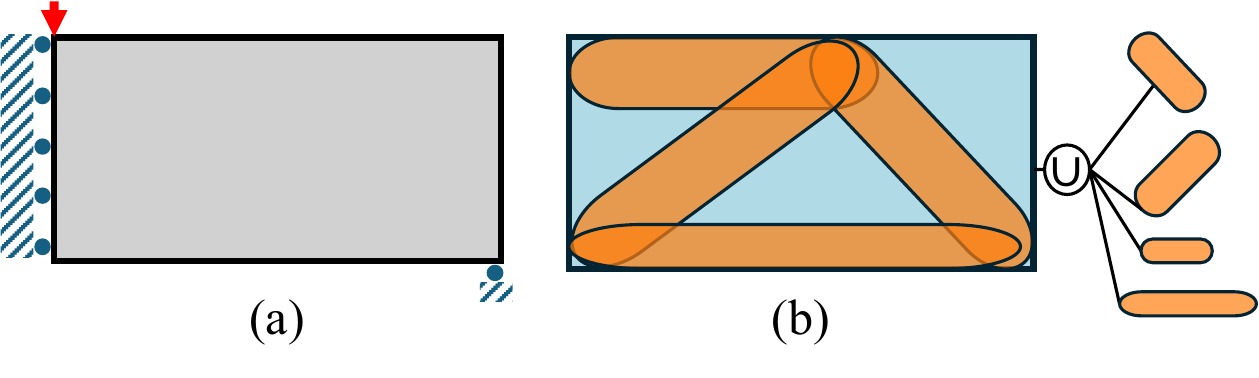}
 		\caption{(a) Classic FMTO parameterizes designs using a union of primitives (such as bars). }
        \label{fig:classic_FMTO}
	\end{center}
 \end{figure}

To address these limitations, alternative techniques, referred to as feature mapping-based topology optimization (FMTO) \cite{wein2020reviewFeatureMapping}, have emerged. These approaches utilize high-level geometric descriptors (or primitives), to represent the design. For the design domain and boundary conditions illustrated in \cref{fig:classic_FMTO}(a), an optimized design obtained using FMTO is presented in \cref{fig:classic_FMTO}(b). By parameterizing the design using primitives, FMTO facilitates the upfront enforcement of manufacturing rules, and also aids in design interpretation \cite{wein2020reviewFeatureMapping}. However, FMTO methods often rely \textit{solely} on Boolean unions, which can limit design flexibility and fail to exploit the complex geometric operations available in modern computer-aided designing (CAD) systems \cite{wein2020reviewFeatureMapping}. While some FMTO methods incorporate Boolean operations other than union, the constructive solid geometry (CSG) tree is typically predefined \cite{zhou2016feature}. The optimization process is then limited to reordering operations or splitting features to introduce new branches \cite{liu2020computer}.

\subsection{Contributions}
\label{sec:intro_contri}

This work introduces an FMTO framework with the following contributions: 
\begin{enumerate}
\item \textbf{Expanded Boolean Operations:} We extend traditional feature-mapping methods, typically limited to Boolean unions, to incorporate subtraction and intersection operations, significantly expanding the design space. Specifically, we employ a unified Boolean operation formulation \cite{liu2024unified}, where an interpolation function represents various Boolean operations.  This formulation enables continuous transitions between different Boolean operations, thus facilitating gradient-based optimization.

\item \textbf{Concurrent Optimization:} Unlike prior research \cite{zhou2016feature} that focuses on optimizing only primitive parameters with fixed operators, our framework optimizes both primitive parameters and the Boolean operators simultaneously.
\end{enumerate}

Since we rely on the unified Boolean operation presented in \cite{liu2024unified}, we also inherit the following features:
\begin{enumerate}
\item \textbf{Mitigating Vanishing Gradients:} The utilization of unified Boolean operations, facilitated by a continuous and monotonic interpolation scheme, enables gradient-based optimization while mitigating the issue of vanishing gradients.

\item \textbf{Primitive Parameterization Flexibility:} The proposed method is agnostic to the specific primitive parameterization, accommodating a wide range of geometric primitives.

\end{enumerate}

\section{Related Work}
\label{sec:litReview}

As mentioned earlier, FMTO parameterizes the topology via high-level geometric features \cite{shapiro2002solid} or primitives that are mapped onto a mesh for analysis \cite{wein2020reviewFeatureMapping, chandrasekhar2023polyto}. FMTO is driven by the need to integrate features into free-form designs, control specific structural dimensions, utilize stock materials, and generate a geometric representation that can be interpreted by CAD systems \cite{wein2020reviewFeatureMapping}. Some of the earliest work in this area combines free-form topology optimization with embedded geometric primitives \cite{qian2004optimal, chen2006parametric, xia2012sensitivity, kang2016structural}; for a detailed review, see \cite{wein2020reviewFeatureMapping, coniglio2020generalized}.

We focus on methods that represent the design exclusively using geometric primitives \cite{norato2018topology, norato2015gpto, chandrasekhar2023polyto, padhy2024photos}. In particular, we discuss four popular approaches that represent the design as a union of primitives capable of translating, scaling, and modifying their shapes \cite{coniglio2020generalized}:

\begin{enumerate}
\item \textbf{Moving morphable components/voids (MMC/MMV) method:} The MMC/MMV method uses geometric primitives, such as B-spline-shaped holes or components, to represent the design \cite{guo2016explicit, zhang2017explicit}. This approach allows for control of the design's geometry by explicitly manipulating the boundaries of these primitives \cite{zhang2017structural}.

\item \textbf{Geometry projection (GP) method:} The GP method \cite{bell2012geometry, norato2015gpto} uses geometric primitives, such as bars \cite{norato2015gpto}, supershapes \cite{norato2018topology, padhy2024tomas} and plates \cite{zhang2016geometry}, to optimize structural designs by projecting these primitives onto a fixed finite element mesh. This approach has been applied in various contexts, including 3D topology optimization \cite{zhang2018geometry}, multi-material design \cite{kazemi2018topology}, and the optimization of unit cells for lattice materials \cite{watts2017geometric}. 

\item \textbf{Method of Moving Morphable Bars (MMB):} The MMB uses round-ended bars as primitives, allowing them to overlap and modify their shape and position within the design \cite{hoang2017topology}.

\item \textbf{Moving Node Approach (MNA):} Finally, the MNA uses polynomial functions to project geometric primitives, representing the building blocks of the design as mass nodes \cite{overvelde2012moving}. 
\end{enumerate}

The above approaches rely on the Boolean union of primitives \cite{coniglio2020generalized}, which can restrict design flexibility.  We propose a generalized framework in (\cref{sec:method}) to enhance design flexibility. \Cref{sec:method} covers the framework's key components: primitive parametrization (\cref{sec:method_primitives}), projection of primitives onto the density field (\cref{sec:method_geometryProjection}), the use of an expanded set of Boolean operations (union, intersection, subtraction) on the density fields (\cref{sec:method_boolean}), and optimization strategy (\cref{sec:method_opt}). Several examples demonstrating the proposed method are presented in \cref{sec:expts}. Finally, \cref{sec:conclusion} concludes this work.

\section{Proposed Method}
\label{sec:method}

\subsection{Overview}
\label{sec:method_overview}
\begin{figure}[H]
 	\begin{center}
		\includegraphics[scale=0.7,trim={0 0 0 0},clip]{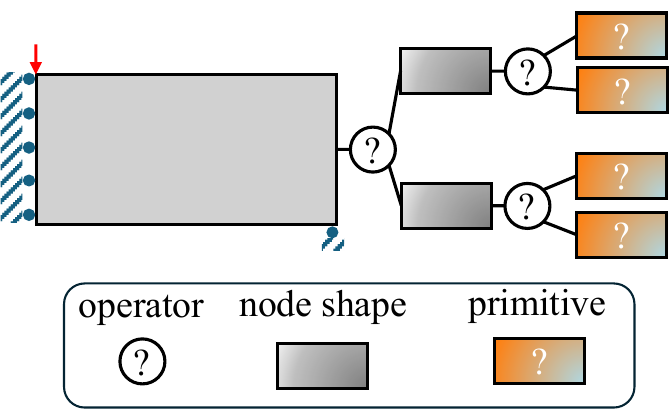}
 		\caption{ Given a design domain with boundary conditions, the proposed method optimizes the boolean operations ($\text{union}(\cup), \text{intersection}(\cap), \text{difference}(\mapsto), \text{and negative difference}(\mapsfrom)$), and the parameters associated with primitives (polygon). The resulting shapes are hierarchically combined till the final design is obtained at the root.}
        \label{fig:variation_optimization_params}
	\end{center}
 \end{figure}

This study focuses on gradient-based topology optimization, which minimizes structural compliance under a volume constraint. We assume that the design domain, loads and restraints, has been prescribed. While the framework is agnostic to the type of primitives used, we will assume that the primitives are polygons for ease of implementation. The CSG tree is assumed to be a perfectly balanced binary tree with a specified depth. Our objective is to obtain an optimal configuration of the polygons at the leaf nodes and Boolean operations at all non-leaf nodes that, when applied, results in the optimized design; see \cref{fig:variation_optimization_params}.

\subsection{Primitive Parametrization}
\label{sec:method_primitives}

\begin{figure}[H]
 	\begin{center}
		\includegraphics[scale=0.7,trim={0 0 0 0},clip]{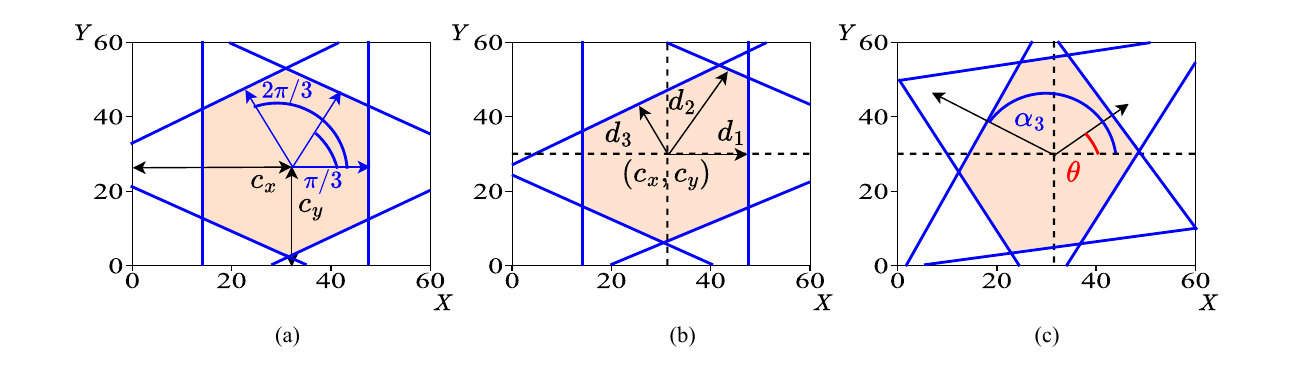}
 		\caption{The polygon's parameters include the (a) center coordinates $(c_x, c_y)$, (b) the distances from the center to the half-spaces $(d_1, \ldots, d_6)$, and (c) an angular offset $\theta$.}
        \label{fig:polygon_primitive}
	\end{center}
 \end{figure}

We populate the leaf nodes of a perfectly balanced binary CSG with \emph{polygonal} primitives. In particular, a tree with a (given) depth of $n_d$ (discussed in \cref{sec:expts}) would correspond to $n_p = 2^{n_d}$ primitives at the leaf nodes. The primitives are parameterized using the method proposed in \cite{deng2020cvxnet, chandrasekhar2023polyto} where:

\begin{enumerate}
    \item Each polygon primitive $p^{(i)}$ is associated with a reference point $(c_x^{(i)}, c_y^{(i)})$, and is the intersection of $S$ ($ S \geq 3)$ (straight-line) half-spaces, resulting in a polygon with $\{3, \ldots, S\}$ sides; $S = 6$ in \cref{fig:polygon_primitive}(a). 

    \item Each half-space $h_j^{(i)}$ is initially oriented at an angle $\bm{\tilde{\alpha}}^{(i)} = \left[0, \pi/3, 2\pi/3, \ldots \right]$ (see \cref{fig:polygon_primitive} (a)) and has an offset distance $d_j^{(i)} > 0$ from the reference point (see \cref{fig:polygon_primitive} (b)).
    \item To vary the overall orientation of the polygon, we allow for the rotation of all of its half-spaces by an angle $\theta^{(i)}$, resulting in the final orientation angle $\alpha_j^{(i)} = \theta^{(i)} + \tilde{\alpha}_j^{(i)}$ for the $j^{\text{th}}$ half-space (see \cref{fig:polygon_primitive}(c)).
\end{enumerate}
In summary, each polygon $p^{(i)}$ is parameterized as:

\begin{equation}
    p^{(i)} = \{c_x^{(i)}, c_y^{(i)}, \theta^{(i)}, d_1^{(i)}, \ldots, d_S^{(i)} \}, \quad  \; i = 1,\ldots, n_p
    \label{eq:param_of_polygon}
\end{equation}

Note that, by construction, each polygon is non-empty and can have between $3$ and $S$ sides \cite{chandrasekhar2023polyto}.

\subsection{Geometry Projection}
\label{sec:method_geometryProjection}

\begin{figure}[h]
 	\begin{center}
		\includegraphics[scale=0.6,trim={0 0 0 0},clip]{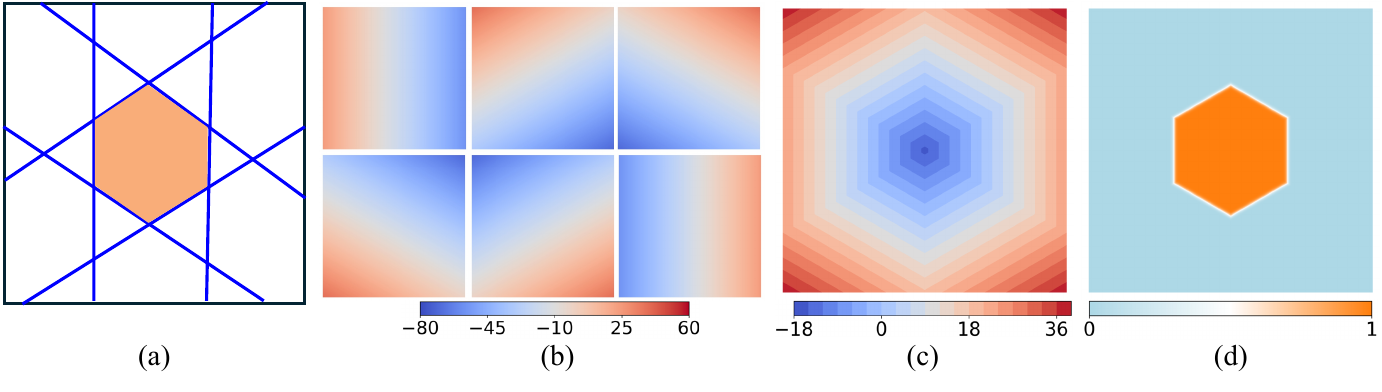}
 		\caption{(a) A polygon, (b) signed distance field (SDF) of half-spaces (c) SDF of polygon, and (c) its projected density.}
        \label{fig:projection}
	\end{center}
 \end{figure}

The aim of geometry projection is to map primitives, defined by the polygon's parameters (\cref{sec:method_primitives}) onto a density field defined over a mesh  \cite{deng2020cvxnet}. To maintain differentiability, we first map each primitive (\cref{fig:projection}(a)) to a signed distance field (SDF), where the value of the SDF at any point ($x, y$) is defined as the shortest signed-distance to the primitive’s boundary (inside being negative and outside being positive). We begin by defining the SDF  of each half-space as (\cref{fig:projection}(b)):

 \begin{equation}
    \hat{\phi}_j^{(i)}(x, y) = (x - c_x^{(i)})\cos(\alpha_j^{(i)}) + (y - c_y^{(i)})\sin(\alpha_j^{(i)}) - d_j^{(i)}
    \label{eq:sdf_hyperplane}
\end{equation}

Since the SDF value of a point inside the polygon with respect to a half-space is negative, the SDF of the entire polygon at a given point is obtained by computing the maximum of the SDFs from each half-space at that point (\cref{fig:projection}(c)). To ensure differentiability, we use the LogSumExp (LSE) approximation of maximum \cite{zhang2021dive}. This yields the SDF of the $i^{th}$ polygon:

\begin{equation}
    \phi^{(i)}(x, y) = \frac{l_0}{t} \log \left( \sum_{j=1}^{S} \exp\left(\frac{t}{l_0} \hat{\phi}_j^{(i)}(x, y)\right) \right)
    \label{eq:sdf_polygon}
\end{equation}
where, $t$ is the LSE scaling factor (discussed later in \cref{sec:expts}) and $l_0$ is the length of the diagonal of the domain bounding box.

Next, we compute the density field of the polygon $\tilde{\rho}(\cdot)$, from $\phi(\cdot)$ using the Sigmoid function; see \cref{fig:projection}(d): 
\begin{equation}
    \tilde{\rho}^{(i)}(x,y) = \frac{1}{1 + \exp(-\frac{\gamma}{l_0} \phi^{(i)}(x,y))}
    \label{eq:density_projection}
\end{equation}
 
Observe that negative SDF values (representing the polygon's interior) are mapped to a density of one (solid), while positive values (representing the polygon's exterior) are mapped to zero (void). SDF values near zero (representing the polygon's boundary), are projected to intermediate density values $\tilde{\rho} \in (0,1)$, with the sharpness of transition controlled by the hyperparameter $\gamma$.

Finally, we impose a threshold filter on $\tilde{\rho}$ to drive intermediate densities towards 1/0 as \cite{wu2017infill}:

\begin{equation}
    \rho(\tilde{\rho}) = \frac{\tanh(\frac{\beta}{2}) + \tanh(\beta (\tilde{\rho} - \frac{1}{2}))}{2 \tanh(\frac{\beta}{2})}
    \label{eq:threshold_filter}
\end{equation}
where the parameter $\beta$ controls the sharpness of the threshold function. The obtained primitive densities are subsequently combined using the Boolean operations (as detailed in the following section) to obtain the design density.

\subsection{Unified Boolean Operations}
\label{sec:method_boolean}

\begin{figure}[H]
 	\begin{center}
		\includegraphics[scale=0.6,trim={0 0 0 0},clip]{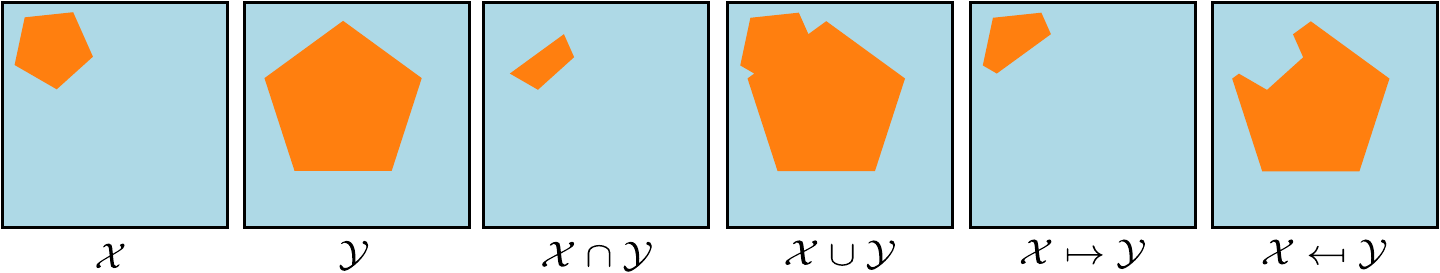}
 		\caption{Boolean operators (union ($\cup$), intersection($\cap$), difference($\mapsto$), negative difference($\mapsfrom$)) defined over the density fields of two pentagonal primitives.}
        \label{fig:boolean_operations}
	\end{center}
 \end{figure}

The final design density is obtained by applying Boolean operators (\cref{fig:boolean_operations}) to the primitive density fields (operands)  defined in the previous section. A successful Boolean operations approach suitable for gradient-based optimization (\cref{sec:method_opt}) should possess the following properties:

\begin{enumerate}
\item Differentiability with respect to operands (density fields).
\item Differentiability with respect to the Boolean operators (union ($\cup$), intersection($\cap$), difference($\mapsto$), negative difference($\mapsfrom$); see \cref{fig:boolean_operations}).
\end{enumerate}

FMTO methods that demonstrate differentiability with respect to the operands \cite{norato2015gpto} have been extensively employed. Further, the union and intersection operators are often approximated using smooth maximum/minimum functions \cite{norato2015gpto, wein2020reviewFeatureMapping, caprinet2022}.  Alternative techniques, such as the smooth blending operator  \cite{ricci1973constructive} and R-functions \cite{rvachev1963analytical, shapiro2007Rfunctions}, have been proposed, exhibiting differentiability with respect to the operands and individual operators.  However, a smooth transition between the operators is difficult to achieve.  Furthermore, while a unified Boolean approach based on the modified R-function \cite{wyvill1999extending} was introduced to facilitate differentiable transitions between operators, it failed to satisfy key Boolean operator axioms, resulting in unexpected behavior \cite{liu2024unified}.

To address these issues, we adopt a unified Boolean operation approach proposed in \cite{liu2024unified}. Let $\mathcal{X}$ and $\mathcal{Y}$ represent two primitives (polygons), with their corresponding density functions $\rho_x$ and $\rho_y$. Then, a generalized Boolean operation between the two primitives is defined as (\cref{fig:boolean_operations}): 

\begin{equation}
    \mathcal{B}(\rho_x, \rho_y ; \bm{b}) = (b_1 + b_2) \rho_x + (b_1 + b_3) \rho_y + (b_0 - b_1 - b_2 - b_3) \rho_x \rho_y
    \label{eq:unified_boolean_operator}
\end{equation}
where $\bm{b} = \{b_0, b_1, b_2, b_3\}$  are interpolating parameters. The interpolating parameters are constrained by $0 \leq b_i \leq 1$ and $\sum\limits_{i=0}^3 b_i = 1$. When $\bm{b}$ is a one-hot vector, we recover the standard Boolean operators (\cref{table:boolean_operations}):

\begin{table}[H]
	\caption{Boolean operations}
	\begin{center}
		\begin{tabular}{  |c|c|c|  }
                \hline
                $\bm{b}$ & $\text{Operation}$ & $\mathcal{B}(\rho_x, \rho_y)$ \\
                \hline 
                $1,0,0,0$ & $\mathcal{X} \cap \mathcal{Y}$ & $\rho_x \rho_y$ \\
                $0,1,0,0$ & $\mathcal{X} \cup \mathcal{Y}$ & $\rho_x + \rho_y - \rho_x\rho_y$ \\
                $0,0,1,0$ & $\mathcal{X} \mapsto \mathcal{Y}$ & $\rho_x - \rho_x\rho_y$ \\
                $0,0,0,1$  & $\mathcal{X} \mapsfrom \mathcal{Y}$ & $\rho_y - \rho_x\rho_y$ \\ \hline
		\end{tabular}
	\end{center}
	
	\label{table:boolean_operations}
\end{table}

\begin{figure}[H]
 	\begin{center}
		\includegraphics[scale=1,trim={0 0 0 0},clip]{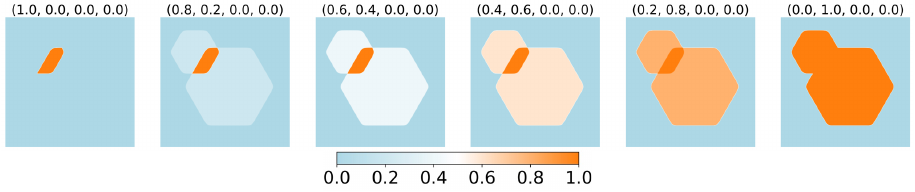}
 		\caption{Continuous interpolation between intersection $(\cap)$ and union $(\cup)$. }
        \label{fig:boolean_operations_interpolation}
	\end{center}
 \end{figure}

\begin{figure}[H]
 	\begin{center}
		\includegraphics[scale=0.4,trim={0 0 0 0},clip]{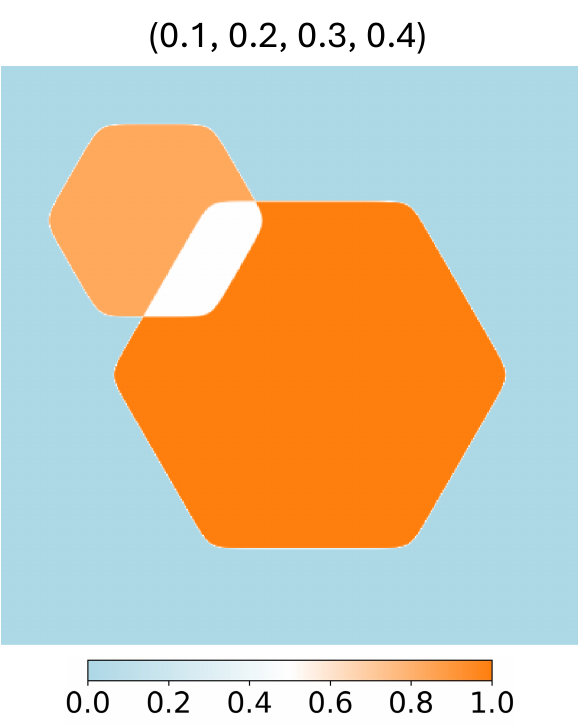}
 		\caption{ The Boolean operand $\bm{b}$ with intermediate non-zero values and the resulting density field.}
        \label{fig:boolean_operations_intermediate}
	\end{center}
 \end{figure}

Observe that \cref{eq:unified_boolean_operator} continuously interpolates between individual operators. For the two polygons in \cref{fig:boolean_operations}, the continuous interpolation between the intersection and union operators is illustrated in \cref{fig:boolean_operations_interpolation}, while a generic operator state is illustrated in \cref{fig:boolean_operations_intermediate}. 

Furthermore,  the function $\mathcal{B}$ is differentiable with respect to both the operands $(\rho_x, \rho_y$) and the operators ($b_i$). For example, the derivative w.r.t operand $\rho_x$ is:

\begin{equation}
    \frac{\partial \mathcal{B}}{\partial \rho_x} = (b_1 + b_2) + (b_0 - b_1 - b_2 - b_3)\rho_y
    \label{eq:deriv_B_rhoX}
\end{equation}
while the derivative with respect to operator coefficient $b_1$ is
\begin{equation}
    \frac{\partial \mathcal{B}}{\partial b_1} = \rho_x + \rho_y - \rho_x\rho_y
    \label{eq:deriv_B_b1}
\end{equation}
 
Finally, with the interpolated operators defined at all non-leaf nodes and the primitive densities at the leaf nodes, we can evaluate the CSG tree bottom-up to obtain the  density field $\rho(x,y)$ of the evolving design at any instance.

\subsection{Finite element analysis}
\label{sec:method_fea}

For finite element analysis, we use here a structured mesh with bilinear quad elements. Having obtained the design density field $\rho(x,y)$, let $\rho_e$ be the density at the center of element $e$.  The corresponding Young's modulus $E_e$ is obtained using the Solid Isotropic Material Penalization (SIMP) model \cite{sigmund2013topology} as:

\begin{equation}
    E_e = E_{min} + (E_0 - E_{min})(\rho_e)^p
    \label{eq:SIMP_material}
\end{equation}
where $E_0$ is Young's modulus of the solid material, $E_{min}$ is a small constant added to prevent a singular global stiffness matrix, and $p$ is the SIMP penalty. One can evaluate the element stiffness matrix as
\begin{equation}
[K_e] = E_e \int\limits_{\Omega_e} [B]^T[D_0][B] d \Omega_e
\label{eq:elem_stiffness}
\end{equation}

where $[B]$ is the strain matrix, and $[D_0]$ is the constitutive matrix with a Young’s modulus of unity and an assumed Poisson's ratio (see Table~\ref{table:defaultParameters}). This is followed by the assembly of the global stiffness matrix $\bm{K}$. Finally, with the imposed nodal force vector $\bm{f}$, we solve the state equation for the nodal displacements $\bm{u} = \bm{K}^{-1}\bm{f}$. 

\subsection{Optimization}
\label{sec:method_opt}

We now summarize various aspects of the optimization problem:

\paragraph{Objective:} We consider here a compliance minimization problem. where the compliance is computed as $J = \bm {f}^T \bm{u}$.

\paragraph{Volume constraint:} The design is subjected to a total volume constraint. With $v_f^*$ being the maximum allowed volume fraction and $v_e$ being the element areas, the volume constraint $g_v$ is defined as:

\begin{equation}
    g_v \equiv \frac{\sum\limits_{e=1}^{N_e} \rho(\bm{x}_e) v_e} {v_f^* \sum\limits_{e=1}^{N_e} v_e} - 1 \leq 0
    \label{eq:vol_cons}
\end{equation}

\paragraph{Optimization variables:}  The optimization variables are as follows:
\begin{enumerate}
    \item Recall that the design is defined by polygon parameters, including center coordinates, angular offset, and plane distances, resulting in ${n_p(S+3)}$ free parameters (see Section~\ref{sec:method_primitives}). Additionally, the design requires $n_b = 2^{n_d} - 1$ boolean operations $(\vec{\bm{b}} = \bm{b}^{(1)}, \ldots, \bm{b}^{(n_b)})$ for a tree of depth ${n_d}$. 
    \item For optimization, we define an augmented normalized design vector $\bm{z} = [\bm{z}_{c_x}, \bm{z}_{c_y}, \bm{z}_{\theta}, \bm{z}_{d}, \bm{z}_b]$ that lie in $[0,1]$. With $\underline{c}_x, \underline{c}_y, \underline{\theta}, \underline{d}$ being the lower bound and $\overline{c}_x, \overline{c}_y, \overline{\theta}, \overline{d}$ being the upper bound on the parameters of the polygons, we can retrieve the unnormalized x-center as $\bm{c}_x \leftarrow \underline{c}_x + (\overline{c}_x - \underline{c}_x) \bm{z}_{c_x}$ and so on. The design variable $\bm{z}_b^{(i)}$, corresponding to the $i^{th}$ boolean operator is transformed into one-hot encoding using a softmax function \cite{liu2024unified}.
    \item All design variables $\bm z$ are uniform-randomly initialized (using \texttt{numpy}) with a default seed value of 2. 

\end{enumerate}

\paragraph{TO problem:} The final TO problem is posed as:  

\begin{subequations}
	\label{eq:optimization_eq}
	\begin{align}
		& \underset{\bm{z}}{\text{minimize}}
		& &J = \bm{f}^T \bm{u}(\bm{z}) \label{eq:opt_objective}\\
		& \text{subject to}
		& & \bm{K}(\bm{z})\bm{u} = \bm{f}\label{eq:opt_fea}\\
		& & & g_{v} (\bm{z}) \leq 0 \label{eq:opt_volCons}\\
            & & &  0 \leq z_i \leq 1 \quad \forall i \label{eq:opt_boxCons}
	\end{align}
\end{subequations}

\paragraph{Optimization method:}  We employ the method of moving asymptotes (MMA) \cite{svanberg1987MMA} to perform the design updates. Specifically, we use a Python implementation with all default parameters corresponding to the version of MMA presented in \cite{svanberg2007mma}. The choices of the MMA move limit, the step tolerance, the Karush-Kuhn-Tucker (KKT) tolerances, etc are specified later under numerical experiments.

\subsection{Sensitivity Analysis}
\label{sec:sensitivity_analysis}

\begin{figure}[]
 	\begin{center}
		\includegraphics[scale=1.2,trim={10 0 50 0},clip]{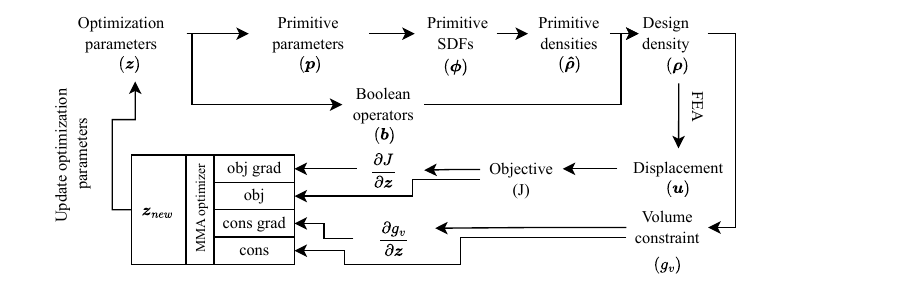}
 		\caption{Optimization loop of the proposed framework.}
 		\label{fig:flowchart}
	\end{center}
 \end{figure}

Sensitivity computation, a crucial aspect of gradient-based optimization, involves determining the derivatives of the objective function and constraints with respect to the optimization parameters. Traditionally, this is conducted manually, which can be labor-intensive and error-prone. However, by utilizing the automatic differentiation (AD) \cite{chandrasekhar2021auto, ian2020AD} capabilities of frameworks such as JAX \cite{jax2018github}, this step can be fully automated, ensuring accurate and efficient computation. In practical terms, we only need to define the forward expressions, and the derivatives of the objective and volume constraint with respect to the optimization variables is automatically computed with machine precision. Finally, we summarize the  algorithm for the proposed framework in \cref{fig:flowchart}.

\section{Numerical Experiments}
\label{sec:expts}

In this section, we conduct several experiments to illustrate the proposed framework. The default parameters in the experiments are summarized in Table~\ref{table:defaultParameters}. All experiments are conducted on a MacBook M2 Air, using the JAX library \cite{jax2018github} in Python.  Additionally, all design parameters $\bm z$ are uniform randomly initialized.

\begin{table}[]
	\caption{Default parameters}
	\begin{center}
		\begin{tabular}{  r | p{62mm}  }
			
        Parameter & Description and default value \\ \hline
        $E_0$, $\nu$ & $E_0 = 1$ and Poisson's ratio $\nu = 0.3$ \\
   
        $n_e$ & Number of FEA mesh elements of ($60 \times 30$) \\

        $l_x$, $l_y$ & Length of domain (60, 30) along $x$ and $y$ \\

        $S$ & Sides of polygon = 6 \\

        $(\underline{c}_x$, $\overline{c}_x)$ & Polygon x-center range $(0.05l_x, 0.95l_x)$ \\

        $(\underline{c}_y$, $\overline{c}_y)$ & Polygon y-center range $(0.05l_y, 0.95l_y)$ \\

        $\underline{d}$, $\overline{d}$ & Polygon face offset range $(0, 0.25l_x)$ \\

        $\gamma$ & Sharpness param in density projection = 100 \\

        $\beta$ & Sharpness param in threshold filter = 8 \\

        $t$ & Scaling factor of the LSE function = 100 \\

        seed & Random seed for initialization = 2 \\

        maxIter & Max number of optim iters = $200$\\
        
        move limit & MMA  step size = $0.05$\\
   
        kkt tolerance & MMA convergence criteria = $10^{-3}$ \\
                
        step tolerance & MMA convergence criteria = $10^{-3}$ \\
			
		\end{tabular}
	\end{center}
	
	\label{table:defaultParameters}
\end{table}

\subsection{Validation}
\label{sec:expts_validation}

\begin{figure}
 	\begin{center}
        \includegraphics[scale=0.5,trim={0 0 0 0},clip]{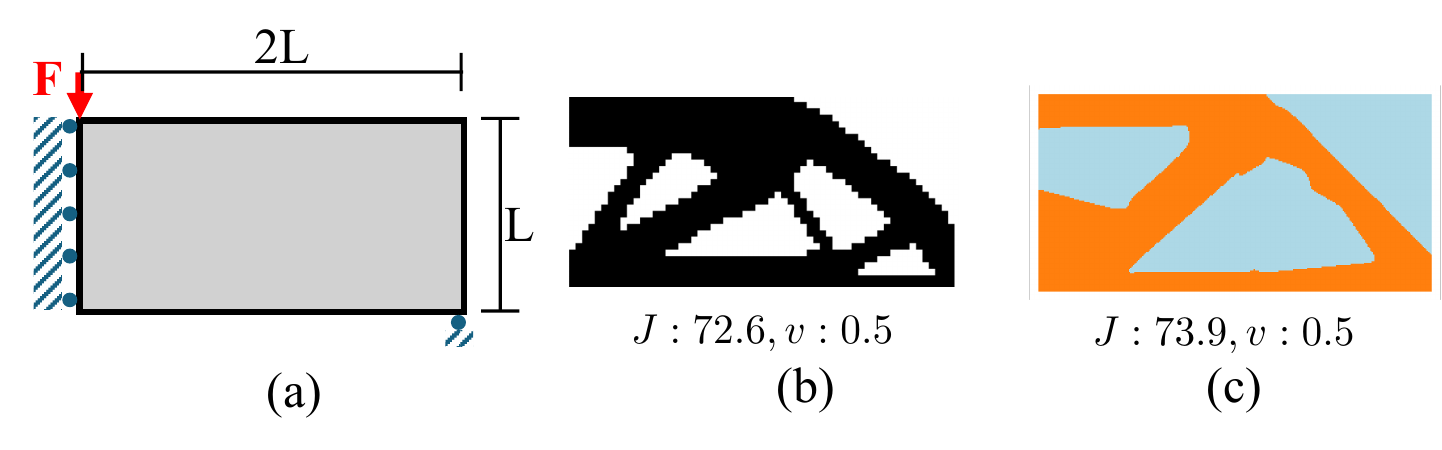}
 		\caption{Validation: (a) MBB problem, (b) SIMP generated topology, and (c) design from the proposed framework.}
        \label{fig:MBB_val}
	\end{center}
 \end{figure}

\begin{figure}
\begin{center}
    \includegraphics[scale=0.6,trim={0 0.6cm 0 0},clip]{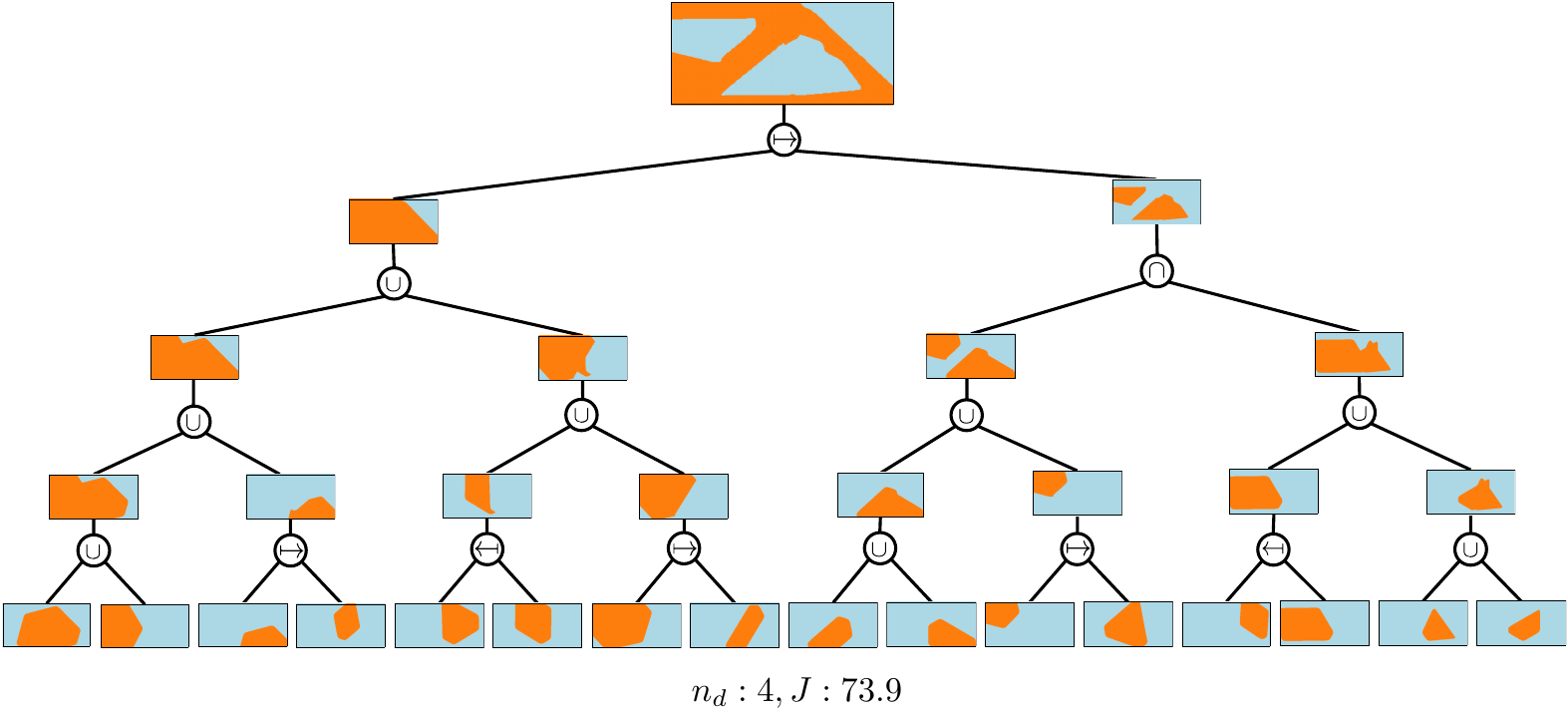}
    \caption{CSG tree with $n_d = 4$. $J = 72.3$ with $v_f^* = 0.5$.}
    \label{fig:mbb_polygon_16}
\end{center}
\end{figure}

We compare the designs from our proposed framework with those obtained using the SIMP-based optimization \cite{andreassen2011efficient}. Using the default parameters outlined in \cref{table:defaultParameters}, and with $n_d=4$, we optimize both the Boolean operators and polygon parameters for an MBB beam (\cref{fig:MBB_val}(a).  In the SIMP implementation, the density field is optimized with a filtering radius of $1.3$, targeting a volume fraction of $v_f^* = 0.5$ ( \cref{fig:MBB_val}(b)). For the proposed method, we display the final design in \cref{fig:MBB_val}(c).  We observe that both methods yield similar designs and performance.  For illustration, we present the complete resulting CSG tree in \cref{fig:mbb_polygon_16}.

\begin{figure}[H]
 	\begin{center}
		\includegraphics[scale=0.35,trim={20 40 0 20},clip]{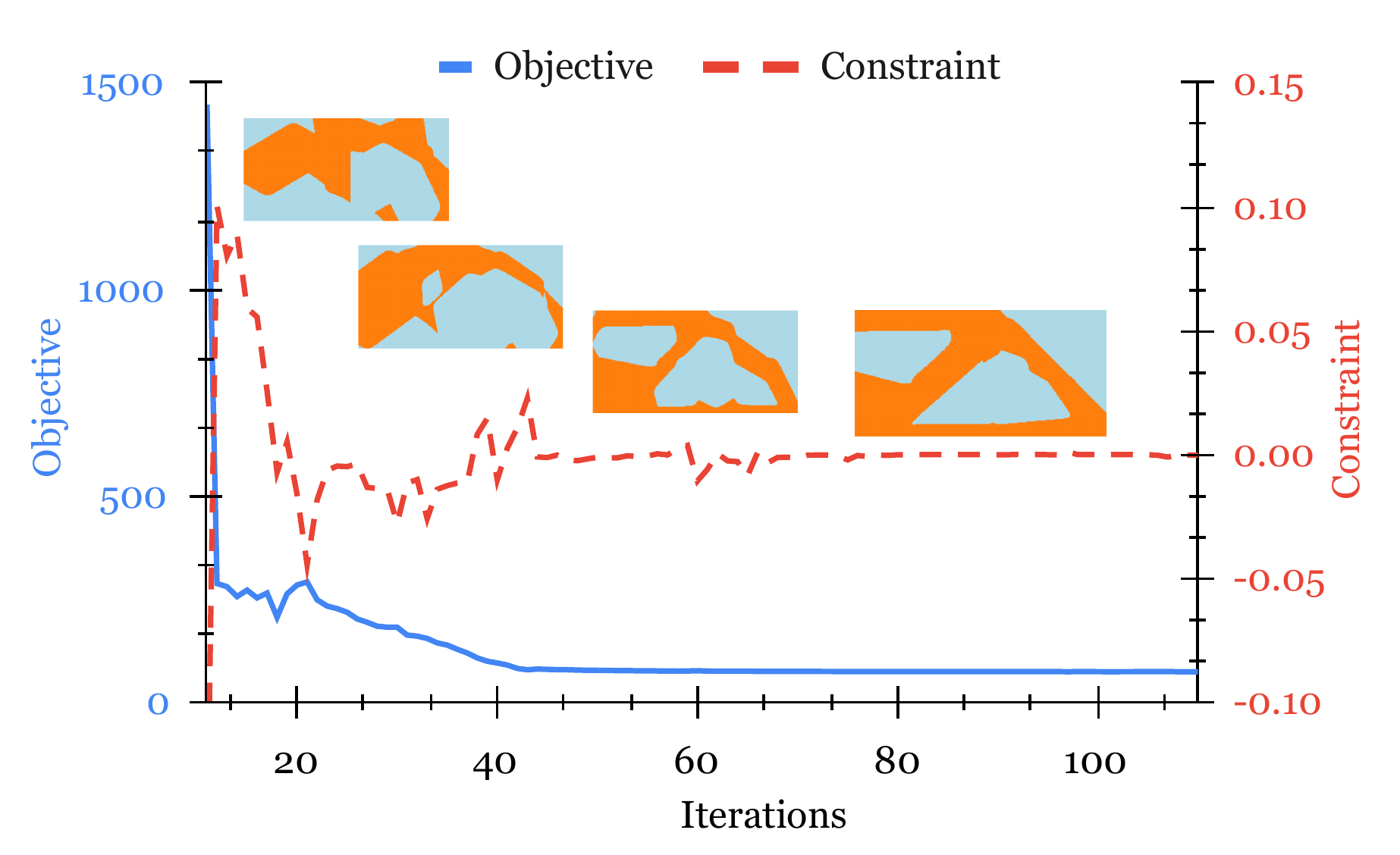}
 		\caption{Convergence of the objective and constraint for an MBB beam with $v_f^* = 0.5$.}
        \label{fig:convergence}
	\end{center}
 \end{figure}

Additionally, the convergence is illustrated in \cref{fig:convergence}. The resulting density fields at the root nodes for the $10^{th}$, $50^{th}$, $75^{th}$, and final ($111^{th}$) iterations are shown as insets. In comparison, the SIMP-based implementation took $126$ iterations to converge. In our framework, the percentage of computational time is as follows: $0.6\%$ for geometry projection, $8.1\%$ for CSG tree evaluation, and $91.3\%$ for FEA and sensitivity analysis; each iteration takes $1.6$ seconds.

\subsection{Effect of Tree Depth}
\label{sec:expts_depth}
\begin{figure}[H]
 	\begin{center}
		\includegraphics[scale=0.5,trim={0 0 0 0},clip]{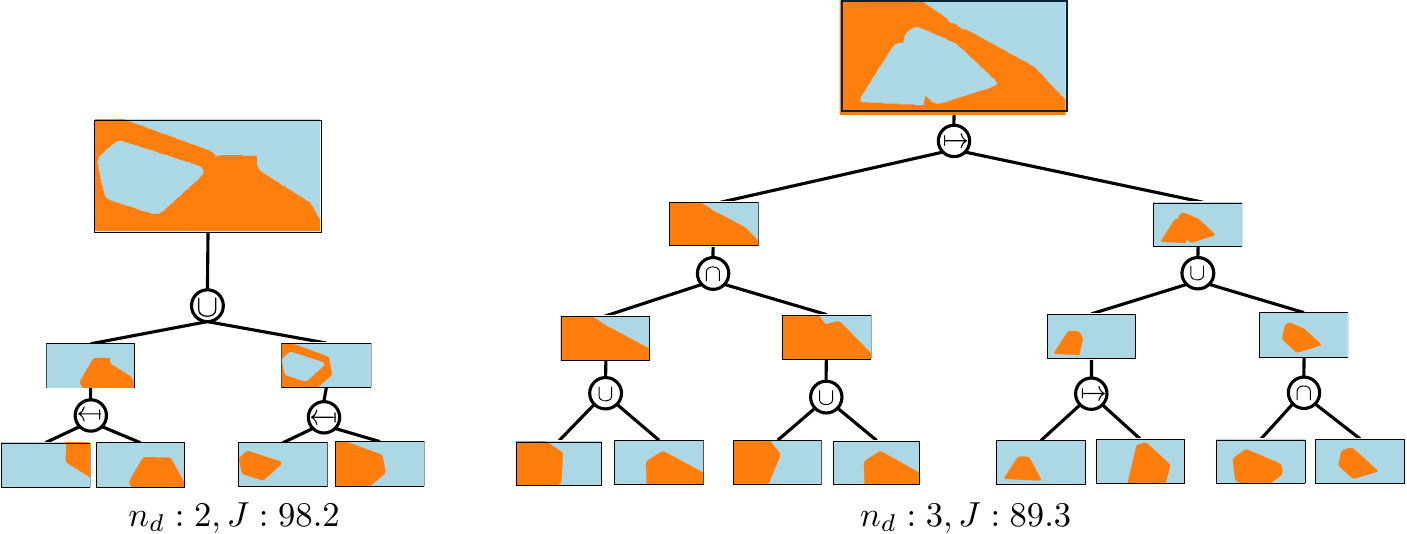}
 		\caption{CSG tree with $n_d = 2$ and $n_d = 3$ with $v_f^* = 0.5$.}
        \label{fig:mbb_polygon_nd_2_3}
	\end{center}
 \end{figure}

\begin{figure}[H]
 	\begin{center}
		\includegraphics[scale=0.6,trim={0 0 0 0},clip]{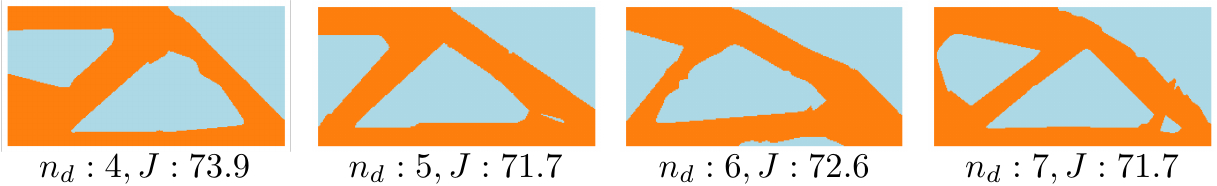}
 		\caption{The final designs and compliances with $n_d=4, 5,6,7$.}
        \label{fig:tree_depth_5_6_7}
	\end{center}
 \end{figure}

In this experiment, we examine the effect of the CSG tree depth $n_d$ on the performance of the design. We revisit the MBB beam in \cref{fig:MBB_val}(a), with $v_f^* = 0.5$, and investigate the impact of different values of $n_d$. The CSG tree for  $n_d =2$ and $n_d =3$ are illustrated in \cref{fig:mbb_polygon_nd_2_3}, while the final designs for $n_d =4,5,6,7$  are illustrated in \cref{fig:tree_depth_5_6_7}. The performance improves as $n_d$ increases; no significant improvements were observed beyond the depth of $n_d = 4$. Based on similar experiments, we recommend using a depth $ n_d \ge 4$; we use a depth of $n_d = 6$ for the remainder of the experiments.

\subsection{Flexibility of Framework}
\label{sec:expts_operators}

\begin{figure}[H]
 	\begin{center}
            \hspace*{-1cm} 
		\includegraphics[scale=0.7,trim={0 2cm 0 0},clip]{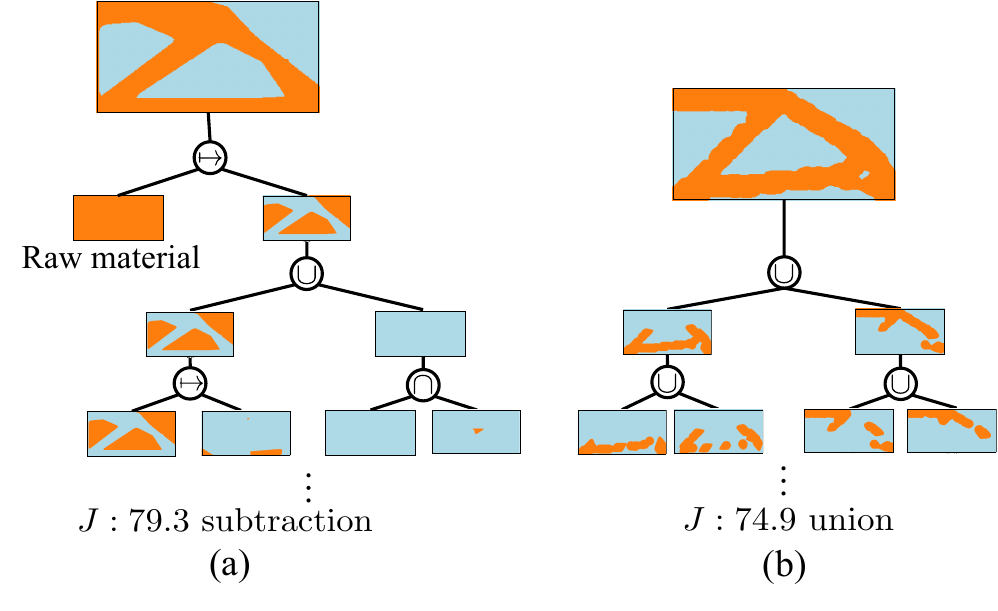}
 		\caption{(a) Difference operator at the root, $J = 79.3$.  (b) Only unions operators allowed, $J = 74.9$.}
        \label{fig:operation_var}
	\end{center}
 \end{figure}

In this experiment, we demonstrate the framework's ability to obtain optimal designs with specific Boolean structures. We again consider the MBB problem (\cref{fig:MBB_val}(a)) with $v_f^* = 0.5$. First, we set the root node operator as a difference operator, resulting in the design shown in \cref{fig:operation_var}(a). Observe that the optimization can sometimes result in empty nodes. These nodes and their descendants are detected and pruned  \cite{tilove1984Pruning}. Next, we only allowed union operators; the resulting topology is illustrated in \cref{fig:operation_var}(b). 

\subsection{Mesh dependency}
\label{sec:expts_mesh}

\begin{figure}[H]
 	\begin{center}
		\includegraphics[scale=0.5,trim={0 0.6cm 0 0},clip]{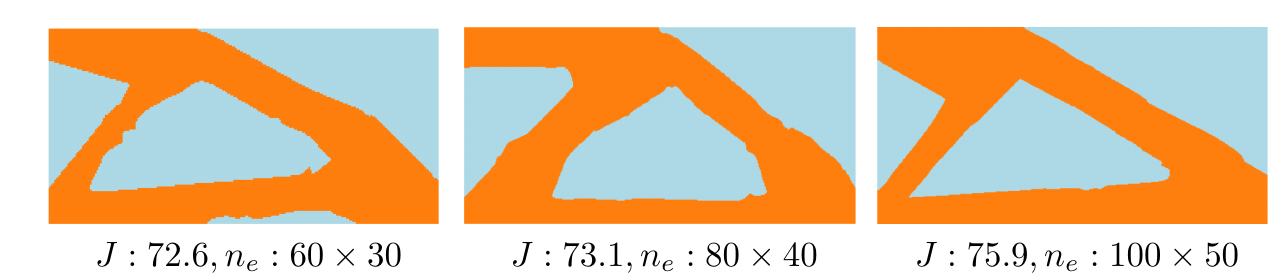}
 		\caption{Dependency on mesh size: (a) $60 \times 30$, $J = 72.6$. (b) $80 \times 40$, $J = 73.1$. (c) $100 \times 50$, $J = 75.9$.}
        \label{fig:mesh_dependency}
	\end{center}
 \end{figure}

Next, we study the effect of the FE mesh on the computed topology using the MBB problem with $v_f^*=0.5$ (\cref{fig:MBB_val}(a)), while keeping all other parameters at their default values. \Cref{fig:mesh_dependency} presents the topologies for varying mesh sizes: $60 \times 30$, $80 \times 40$, and $100 \times 50$ elements. No significant difference in performance was observed across these mesh sizes. However, we note that the designs exhibit less noise as the number of mesh elements increases. The boundary exhibits small features for a coarse mesh, since the latter fails to capture the impact of small features on the performance. However, we have observed that these undesirable features tend to reduce as the mesh size increases.

\subsection{Effect of Initialization}
\label{sec:expts_initDesign}

\begin{figure}[H]
 	\begin{center}
		\includegraphics[scale=0.45,trim={0 0 0 0},clip]{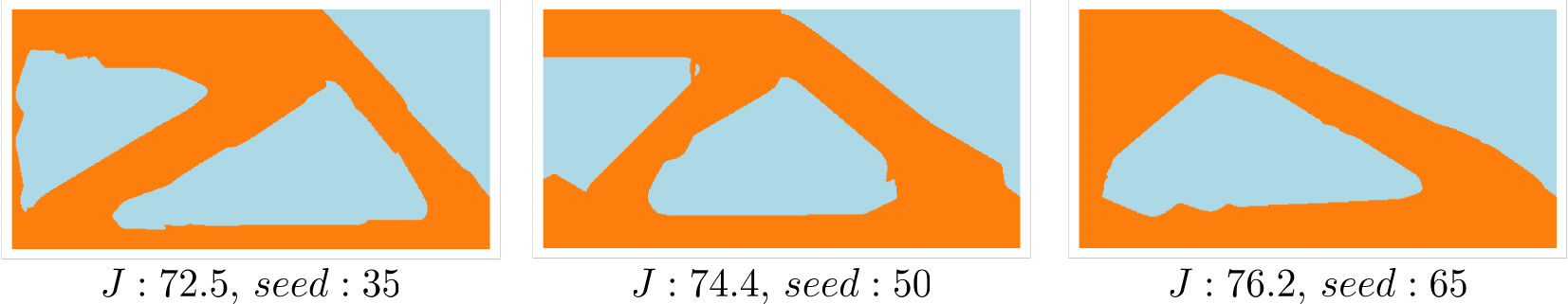}
 		\caption{Dependency on initialization.}
        \label{fig:initialization_analysis}
	\end{center}
 \end{figure}

Recall all design variables $\bm z$ are uniform-randomly initialized with a default seed value of 2. We now investigate the influence of the seed value on the optimal designs. While all topology optimization techniques are inherently sensitive to the initial design \cite{yan2018nonConvexTODensity}, it has been observed that this dependency is particularly pronounced in feature-mapping techniques \cite{zhang2016geometry}. In this specific example, we consider the design of an MBB beam (\cref{fig:MBB_val}(a)) with $100 \times 50$ mesh elements.  The resulting designs and their corresponding performances for various initial seeds are compared in \cref{fig:initialization_analysis}. Observe that while we obtain diverse designs, the performances are similar. This suggests that, as expected, the landscape is highly non-convex with numerous local solutions.

\subsection{Pareto}
\label{sec:expts_pareto}

\begin{figure}[H]
 	\begin{center}
		\includegraphics[scale=0.4,trim={0 0 0 0},clip]{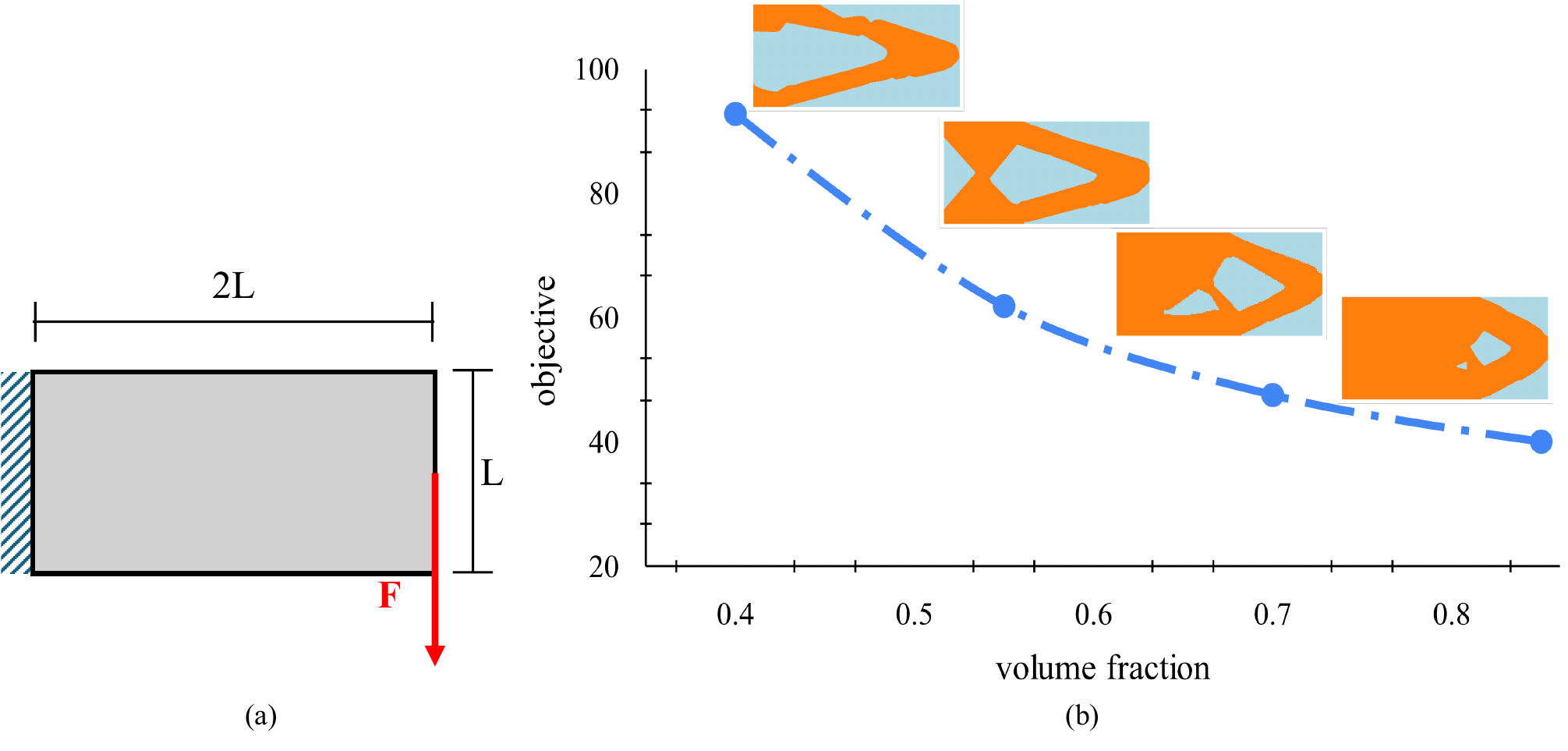}
 		\caption{Trade-off between compliance and volume.}
        \label{fig:pareto_tradeoff}
	\end{center}
 \end{figure}

An essential consideration during the design phase is the exploration of the Pareto front; evaluating the trade-offs associated with various design choices. Let us consider the mid cantilever beam problem illustrated in \cref{fig:pareto_tradeoff}(a). With $100 \times 50$ mesh elements,  $n_d = 6$ and $S = 6$, we investigate the trade-off between the structure's compliance and volume fraction. \cref{fig:pareto_tradeoff}(b) illustrates the resulting designs (and the intermediate densities at the first depth) at various volume fractions. As anticipated, we observe an increase in compliance as the allowable volume fraction is decreased. We did not impose explicit symmetry requirements on the CSG tree.

\section{Conclusion}
\label{sec:conclusion}

We introduce an FMTO framework that concurrently optimizes both the primitive parameters and the Boolean operators (union, intersection, and subtraction). Notably, by leveraging a unified Boolean operation approach, the framrwork improves upon existing FMTO methods that typically rely on predefined CSG trees and/or are limited to optimizing only primitive parameters operated with unions. The efficacy of the proposed method was demonstrated through various numerical examples.

The generated tree structure appears unnatural and dissimilar to human-designed trees. This presents an opportunity to combine the proposed method with machine learning-based frameworks \cite{Koch_2019ABCDataset, li2023secad} to mimic human-designed CSG trees. The method also suffers from some of the shortcomings inherent to FMTO methods. For example, the optimizer is highly susceptible to being trapped in local optima; it is also sensitive to initialization. Additionally, the optimizer may fail to converge to an optimal solution when fewer primitives are used \cite{wein2020reviewFeatureMapping}.

Nevertheless, several avenues for future research remain. For instance, manufacturing constraints such as symmetry, feature size control must be incorporated.  One promising direction is to incorporate CNC operations into the TO process, ensuring that designs are both optimal and manufacturable \cite{yavartanoo2024cnc}. Further, optimizing the primitive type and depth of the tree in conjunction with the operators and primitive parameters requires investigation \cite{caprinet2022, yu2024d2csg}.

\section*{Acknowledgments}
The University of Wisconsin, Madison Graduate School supported this work. 

\section*{Compliance with ethical standards}
The authors declare that they have no conflict of interest.

\section*{Replication of Results}
The Python code is available at \href{https://github.com/UW-ERSL/TreeTOp}{github.com//UW-ERSL/TreeTOp}

\bibliographystyle{unsrt}  
\bibliography{references}

\end{document}